\documentclass[12pt]{article}

\usepackage[intlimits]{amsmath}
\usepackage[cp1251]{inputenc}
\usepackage[T2A]{fontenc}
\usepackage[russian,english]{babel}
\DeclareMathOperator{\tr}{tr}
\usepackage[dviwin]{graphicx}
\usepackage{amsfonts}

\usepackage{feynmf}
\usepackage{geometry}

\begin{document}

\title{The condensate $\langle \tr(A_{\mu}^{2}) \rangle$ in commutative \\ and noncommutative theories}
\author{R.N.Baranov$^{\ddag}$, D.V.Bykov$^{\ddag}$ , A.A.Slavnov$^{\ddag,\natural}$}
\date{}
\maketitle

\language 4

$\ddag$ M.V.Lomonosov Moscow State University, Physics Faculty,
Leninskie gory, bld. 1-2, GSP-2, 119992 Moscow, Russia

$\natural$ V.A.Steklov Mathematical Institute, Gubkina str., bld.
8, GSP-1, 117966 Moscow, Russia
\newline
\begin{center}
\textbf{Abstract}
\end{center}

It is shown that gauge invariance of the operator $\int \, {\rm
dx} \, \tr(A_{\mu}^{2}-\frac{2}{g \xi} x^{\nu} \theta_{\mu\nu}
A^{\mu})$ in noncommutative gauge theory does not lead to gauge
independence of its vacuum condensate. Generalized Ward identities
are obtained for Green's functions involving operator
$\underset{\Omega \rightarrow \infty}{{\rm lim}}\frac{1}{\Omega}
\int\limits_{\Omega} \, {\rm dx} \, \tr(A_{\mu}^{2})$ in
noncommutative and commutative gauge theories.

\section{Introduction}

Those vacuum condensates, which are order parameters at phase
transitions, play a special role in quantum field theory. Examples
are provided by the Higgs field condensate ($\langle \phi \rangle
\neq 0$ - signals the phase of spontaneously broken symmetry) and
the chiral order parameter ($\langle \bar{q} q \rangle \neq 0$ -
signals the phase of broken chiral symmetry). Arguments in favor
of the fact that the condensate of canonical (mass) dimension two
$\langle \tr A_{\mu}^{2} \rangle$ may be useful in the study of
the confinement-deconfinement phase transition in QCD (here
$A_{\mu}$ are the gluon fields) have been presented in
~\cite{zakharov}. The operator $\tr A_{\mu}^{2}$ is not gauge
invariant, but that does not mean that all its matrix elements are
not invariant either. For instance, consider an operator
$F_{\mu\nu}^{2} + n_{\mu}A_{\mu}$ in $U(1)$-theory, where
$n_{\mu}$ is a constant vector. It is clear that this operator is
not gauge invariant, but its v.e.v. coincides with $\langle \tr
F_{\mu\nu}^{2}\rangle$ and is invariant. In the papers
~\cite{slavnov} it was shown that in noncommutative field theory
the condensate $\langle \tr A_{\mu}^{2} \rangle$ does not change
under gauge transformations with a gauge function $\omega(x)$,
which approaches unity at infinity fast enough. On the basis of
some conjectures in ~\cite{slavnov} it was concluded that the
condensate $\langle \tr(A_{\mu}^{2}) \rangle$ does not depend on
the gauge fixing parameter, both in noncommutative, as well as in
commutative theory. In this work it will be shown that the
invariance of the theory with respect to gauge transformations
with functions $\omega (x)$, which approach unity at infinity fast
enough, does not imply gauge independence of the condensate
$\langle \tr(A_{\mu}^{2}) \rangle$. Noninvariance of the
condensate provides obvious restrictions for the possibility of
using it as an order parameter: for this purpose it is necessary
that the collection of values of the coupling constant, at which
the order parameter vanishes, is invariant under gauge
transformations. From this point of view it seems more convenient
to use not the condensate, but its minimal value with respect to
gauge transformations: $L \overset{{\rm def}}{=}
\underset{\omega}{{\rm min}} \, \langle \tr
(A_{\mu}^{\omega})^{2}\rangle$, as it was suggested in
~\cite{zakharov}.

\section{Gauge invariance in classical and quantum \\noncommutative theory}

In this section it will be shown that the invariance of some
quantity under gauge transformations with a gauge function, which
approaches unity fast enough at infinity, does not guarantee gauge
independence of all its matrix elements in quantum theory. Let us
discuss a concrete example, which illustrates this statement and
which is at the same time closely related to the condensate
$\langle \tr A_{\mu}^{2} \rangle$.

We will remind the main steps of the proof ~\cite{slavnov} that
the quantity $\int \,{\rm dx} \,{\rm tr}( A_{\mu}^{2}-\frac{2}{g
\xi} x^{\nu} \theta_{\mu\nu}A^{\mu})$ is gauge invariant in
classical $U(N)$ theory. The easiest way to achieve the desired
result is to start with a matrix model. Its dynamical variable is
the field
\begin{equation}\label{1}
B_{\mu} = A_{\mu} + \frac{1}{\xi} \theta^{-1}_{\mu\nu}x^{\nu}.
\end{equation}
If $A_{\mu}$ transforms as a noncommutative Yang-Mills field, the
gauge transformations for $B$ have a simple form ~\cite{nekrasov}:
\begin{equation}\label{2}
B^{\omega}_{\mu} = \omega \star \, B_{\mu} \star \omega^{-1},
\end{equation}
where $'\star'$ stands for noncommutative Moyal product. It is
easy to build a gauge-invariant quantity $\int \,{\rm dx}\,{\rm
tr}(B_{\mu}^{2})$ and make a shift (\ref{1}). In this way we
obtain a gauge-invariant term $\int \,{\rm dx} \,{\rm
tr}(A_{\mu}^{2}-\frac{2}{g \xi} x^{\nu} \theta_{\mu\nu}A^{\mu})$.

Let us, however, mark a subtlety in the above-mentioned proof: we
used the relation $\int \,{\rm dx}\,{\rm tr} (A \star B - B \star
A)=0$, which is only true if $A$ and $B$ satisfy certain
asymptotic falloff conditions as $\underset{\mu}{{\rm
min}}\,|x_{\mu}|\, \rightarrow\, \infty$. In a proof of classical
gauge invariance the function $\omega (x)$, indeed, can be chosen
to approach unity as fast as we like, but in a proof of the
independence of matrix elements in quantum theory it is not always
possible.

For simplicity we set $N=1$, i.e. consider the gauge group $U(1)$.
Let us find the variation of the v.e.v. $\langle C \rangle
\overset{def}{=} \underset{\Omega \rightarrow \infty}{{\rm lim}}
\, \int\limits_{\Omega} \,{\rm dx}\,{\rm tr}(\langle
A_{\mu}^{2}-\frac{2}{g \xi} x^{\nu}
\theta_{\mu\nu}A^{\mu}\rangle)$ under the transformation $\delta
A_{\mu} = \frac{\delta \alpha}{2 \alpha} D_{\mu}\,\int \,
M^{-1}(x,y) \,
\partial A(y) \, {\rm dy}$, which corresponds to the transition from the $\alpha$ to the
$\alpha + \delta \alpha$ gauge, to zeroth order in the coupling
constant:
\begin{equation}\label{3}
\delta \langle C \rangle = \frac{\delta \alpha}{\alpha} \langle
\int \,{\rm dx}\,{\rm tr}(A_{\mu}\, \partial^{\mu}
\beta_{0})\rangle - \frac{\delta \alpha}{g \xi \alpha} \langle
\int \,{\rm dx}\,{\rm tr} (x^{\nu} \theta_{\mu\nu} (\partial^{\mu}
\beta + g[A^{\mu},\beta_{0}]))\rangle ,
\end{equation}
where $\beta = \int M^{-1}(x,y) \,
\partial A(y) {\rm dy};\; \beta_{0}=\beta|_{g=0}$. We will consider each term separately.
First of all,
\begin{equation}\label{4}
\langle \int \,{\rm dx}\,{\rm tr}(A_{\mu}\, \partial^{\mu}
\beta_{0})\rangle = - \int \, {\rm dy}\, D_{c}(x-y) \underset{= -
\alpha \delta(x-y)}{\underbrace{\langle\partial A(x)\,
\partial A(y)}}\rangle = \alpha D_{c}(0)
\end{equation}
Hereafter the horizontal bracket indicates that we are using the
Ward identity for the two-point function.

The second term $\langle \int \,{\rm dx}\,{\rm tr} (x^{\nu}
\theta_{\mu\nu}
\partial^{\mu} \beta) \rangle$ vanishes, because $$\partial^{\mu}_{x} \langle \beta(x) \rangle =
\partial^{\mu}_{x} \langle \int \, {\rm dy} \, \bar{c}(x) \, c(y)\, \partial A(y)
\rangle= \partial^{\mu}_{x} \,\int \,{\rm dy} \, F(y-x)=0.$$ In
the last equation we used translation invariance of the theory. We
stress that it is translation invariance which makes it impossible
to integrate by parts (with respect to $'x'$). Now we turn to the
last term $\langle \int \, {\rm dx} \, x^{\nu} \theta_{\mu\nu}\,
[A^{\mu},\beta] \rangle$. We write it in the first order in $\xi$
(it is clear that in the zeroth order it is zero, and we can
restrict ourselves to the first order as all relations should hold
in every order):
\begin{equation}\label{5}
\langle \int \, {\rm dx} \, x^{\nu} \theta_{\mu\nu}\,
[A^{\mu},\beta_{0}] \rangle = \frac{\xi}{2} \langle \int \, {\rm
dx} \, x^{\nu} \theta_{\mu\nu} \theta^{\alpha \gamma}
\partial_{\alpha} A^{\mu} \, \partial_{\gamma} \beta_{0}(x)
\rangle
\end{equation}
In classical field theory $\beta_{0}(x)$ is a fast decreasing
function. In this case we would be able to integrate by parts in
the last equality and we would find that the resulting
contribution cancels (\ref{4}). However, in our case
$\beta_{0}=\beta_{0}[A]$ is a concrete function and we cannot
change its asymptotic behavior. Therefore,
\begin{eqnarray}
\nonumber \langle \partial_{\alpha} A^{\mu} \, \partial_{\gamma}
\, \int \, D_{c}(x-y) \, \partial A(y) \,{\rm dy} \rangle & =&
 - \int \, {\rm dy}\, \partial^{y}_{\gamma} D_{c}(x-y)
 \underset{= - \alpha \partial^{x}_{\alpha}\partial^{x}_{\mu}D_{c}(x-y)}{\underbrace{\langle \partial_{\alpha}A^{\mu}(x) \, \partial A(y)\rangle}} = \\
\label{6} & = & - \alpha \int \, {\rm dy} \,
\partial_{\gamma}D_{c}(y)\,
\partial_{\alpha}\partial^{\mu}D_{c}(y) = 0
\end{eqnarray}

Thus, the full variation of $\langle C \rangle$ is
\begin{equation}\label{7}
\delta \langle C \rangle = \delta \alpha \, D_{c}(0) \neq 0,
\end{equation}
unless a specific regularization is chosen, which provides the
equality $D_{c}(0)=0$ (for example, dimensional regularization).

In the work ~\cite{bykov} it was shown that the following relation
holds in the general case:
\begin{equation}\label{7.1}
{\rm \frac{d}{d \alpha}} \langle {\rm tr} (A_{\mu}^{2})
\rangle_{\alpha=0} = \langle {\rm tr} (\bar{c} c)
\rangle_{\alpha=0} ,
\end{equation}
It is consistent with (\ref{7}), because the ghost condensate in
the Lorentz gauge in the zeroth order of perturbation theory is
equal to $D_{c}(0)$.

\section{Generalized Ward identities in the presence \\ of a dimension two
condensate in commutative \\ and noncommutative theory}

In the preceding section we found out that the condensate $\langle
A_{\mu}^{2}\rangle$ depends on the gauge fixing parameter.
Nevertheless, generalized Ward identities for Green's functions
involving operator $C$ hold in the noncommutative theory. It is
due to the fact that both in commutative and noncommutative theory
in order to obtain the generalized Ward identities we make a
transformation $\delta A_{\mu} = D_{\mu} \mathrm{M}^{-1} \phi$
with an arbitrary gauge function $\phi$, and, of course, we can
choose one, which decreases at infinity.

Let us consider the generating functional for the Green's
functions with an insertion of operator $C$:
\begin{equation}\label{8}
\mathcal{Z}_{C}[J] = Z^{-1} \, \int \, C \, \cdot
\exp{\left[i\mathcal{S}+\frac{i}{2\alpha} \int \,{\rm dx}\,
(\partial A)^{2} + i \, \int \, J^{\mu} A_{\mu} \, {\rm
dx}\right]} \, \cdot \, {\rm det} \mathrm{M} \; \mathcal{D}A
\end{equation}
To obtain the generalized Ward identities, as usual, we make a
change of variables $A_{\mu} \rightarrow A_{\mu} + \delta
A_{\mu}$, where
\begin{equation}\label{8.1}
\delta A_{\mu} = D_{\mu} \mathrm{M}^{-1} \phi ,
\end{equation}
where $\phi$ is an arbitrary function. Then we get
\begin{equation}\label{9}
-\frac{i}{\alpha} \partial_{\mu} \left[ \frac{\delta
\mathcal{Z_{C}}}{\delta J_{\mu}} \right] + \int \, {\rm dz} \,
J_{\mu}(z) D_{\mu}^{z}\mathrm{M}^{-1}(z,y) = 0
\end{equation}
This identity coincides with the one that would take place in the
absence of operator $C$ (i.e. for $\mathcal{Z}$ rather than for
$\mathcal{Z}_{C}$), as it should have been expected, because
operator $C$ is invariant under gauge transformations (\ref{8.1})
with a decreasing function $\phi$.

Let us now consider Green's functions with one insertion of
operator $\underset{\Omega \rightarrow \infty}{{\rm
lim}}\frac{1}{\Omega}\int_{\Omega} \, {\rm dx} \,
\tr(A_{\mu}^{2})$ in the commutative theory. They have the
following property:
\begin{equation}\label{15}
\exists \; \underset{\Omega \rightharpoonup \infty}{\mathrm{lim}}
\; \frac{1}{\Omega} \int\limits_{\Omega} {\rm d^{4}x} \, \langle
{\rm [A_{\mu}^{2}(x)]} \cdot \prod\limits_{i}
\mathcal{O}_{i}(y_{i})\rangle = \langle {\rm [A_{\mu}^{2}(0)]}
\rangle \cdot \langle \prod\limits_{i} \mathcal{O}_{i}(y_{i})
\rangle ,
\end{equation}
where $\{\mathcal{O}_{i}: i=1...n\}$ is a set of elementary, i.e.
not composite, operators, and the square brackets indicate that
the operator is renormalized. Later on we will use the notation
$\mathcal{O}(Y) \equiv \prod\limits_{i} \mathcal{O}_{i}(y_{i}) $.

To prove this relation it is sufficient to show that the connected
diagrams corresponding to the l.h.s. of (\ref{15}) vanish in the
limit of infinite volume.

For definiteness we choose $\Omega$ to be a four-dimensional ball:
$\Omega=\mathcal{K}^{4}_{R}$. Next we perform a Fourier
transformation with respect to '$x$' in the l.h.s. of (\ref{15})
and integrate explicitly over '$x$', using the formula
\begin{equation}\label{16}
\int\limits_{\mathcal{K}^{4}_{R}} \, d^{4}x \, e^{ipx} =  \frac{4
\pi^{2} R^{2}}{p^2} \, J_{2}(pR) ,
\end{equation}
where $p\equiv \sqrt{p^{2}}$ and $J_{2}$ is the Bessel function of
the second kind. Thus,
\begin{eqnarray}\label{17}
\underset{\Omega \rightharpoonup \infty}{\mathrm{lim}} \;
\frac{1}{\Omega} \int\limits_{\Omega} {\rm d^{4}x} \, \langle {\rm
[A_{\mu}^{2}(x)]} \cdot \mathcal{O}(Y)\rangle = \underset{R
\rightharpoonup \infty}{\mathrm{lim}} \; \frac{4 \pi^{2}
R^{2}}{\Omega_{\mathcal{K}^{4}_{R}}} \, \int \, d^{4}p \,
\frac{J_{2}(pR)}{p^{2}} \, \langle {\rm [A_{\mu}^{2}(p)]} \cdot
\mathcal{O}(Y)\rangle \\ \nonumber = \underset{R \rightharpoonup
\infty}{\mathrm{lim}} \; \frac{4
\pi^{2}}{\Omega_{\mathcal{K}^{4}_{R}}} \, \int \, d^{4}k \,
\frac{J_{2}(k)}{k^{2}} \, \langle {\rm [A_{\mu}^{2}(\frac{k}{R})]}
\cdot \mathcal{O}(Y)\rangle
\end{eqnarray}
So, we are interested in the asymptotics of the Green's functions
$\langle {\rm [A_{\mu}^{2}(p)]} \cdot \mathcal{O}(Y)\rangle$ as $p
\rightharpoonup 0$. Below we will show that the singularity of the
v.e.v. in this limit is not worse than logarithmic. For that
purpose we will now calculate the infrared index of the diagram.

\par
\begin{figure}
\begin{center}
\begin{fmffile}{diaglarge}
    \begin{fmfgraph*}(300,150)
        \fmfleft{x} \fmfrightn{y}{6}
        \fmfbottomn{b}{4} \fmftopn{t}{4}
        \fmf{photon}{x,b_int1,b_int2,b_int3,r_int1,r_int2}
        \fmf{photon}{x,t_int3,t_int2,t_int1,r_int4,r_int3}
        \fmf{dots}{r_int2,r_int3}
        \fmf{photon}{b2,b_int2} \fmf{photon}{t2,t_int2}
        \fmf{photon,tension=6}{r_int1,y1} \fmf{photon,tension=6}{r_int4,y6}
        \fmf{dots}{y2,y3,y4,y5}
        \fmf{phantom,tension=4}{b1,b_int1} \fmf{phantom,tension=4}{t1,t_int3}
        \fmf{phantom,tension=4}{b3,b_int3} \fmf{phantom,tension=4}{t3,t_int1}
        \fmf{photon}{b_int1,int}
        \fmf{photon}{b_int3,int}
        \fmf{photon}{t_int1,int}
        \fmf{photon}{t_int3,int}
        \fmf{phantom}{r_int1,int} \fmf{phantom}{r_int4,int}
        \fmfv{decor.shape=circle,decor.filled=0,decor.size=20mm}{int}
        \fmfdot{x,b_int1,b_int2,b_int3,r_int1,r_int4,t_int1,t_int2,t_int3}
        \fmf{phantom_arrow,label=$k$,label.side=right}{x,b_int1} \fmf{phantom_arrow,label=$k$}{t_int3,x}
        \fmf{phantom_arrow,label=$k+l_1$,label.side=right}{b_int1,b_int2} \fmf{phantom}{t_int3,t_int2}
        \fmf{phantom_arrow,label=$p_1$,label.side=right}{b2,b_int2} \fmf{phantom_arrow,label=$-p_n$}{t_int2,t2}
        \fmf{phantom_arrow,label=$k+l_1+p_1$,label.side=left}{b_int2,b_int3} \fmf{phantom_arrow}{t_int1,t_int2}
        \fmf{phantom_arrow,label=$k+L+p_1$,label.side=right,tension=4}{b_int3,r_int1}
          \fmf{phantom,tension=4}{r_int4,t_int1}
        \fmf{phantom_arrow,label=$l_1$}{int,b_int1} \fmf{phantom}{int,t_int3}
        \fmf{phantom_arrow,label=$l_2$,label.side=left}{int,b_int3} \fmf{phantom}{int,t_int1}
    \end{fmfgraph*}
\end{fmffile}
\end{center}
\caption{Schematic representation of a connected 1PI diagram. At
nonexceptional external momenta the propagators $(k+l_1+p_1),
(k+L+p_1),\ldots ,(k-p_n)$ do not contribute to $\omega_{ir}$.
Here we have used the notation $L=l_1+l_2$.} \label{fig_large}
\end{figure}
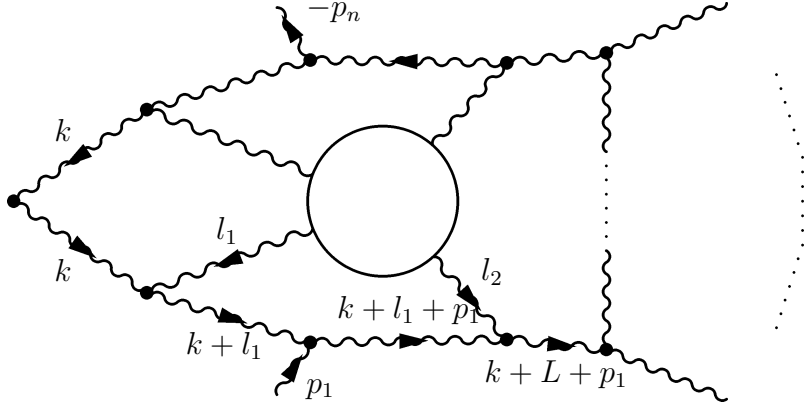

 Consider ~(Fig.~\ref{fig_large}) an arbitrary one-particle-irreducible diagram
with nonexceptional external momenta $p_1\, , \ldots \, ,$ $p_n\,
; \; p_1+\ldots+p_n=0$. We denote by $G_1,\ldots,G_k$ the
diverging subgraphs of the diagram $G$. The Bogoliubov-Parasyuk
$R$-operation acts in the following way:
\begin{equation}\label{R_operation}
    R(G) = \vdots [1-M(G_1)]\ldots[1-M(G_k)] \vdots
\end{equation}
According to the R-operation (\ref{R_operation}), the renormalized
diagram represents a sum of the unrenormalized diagram and the
diagrams which are obtained from the initial one by the
substitution of coefficient functions ~$F_i$ of some divergent
subgraphs by the first terms of their Taylor expansion with
respect to the momenta, which are external for these subgraphs, at
some point ~$\widetilde{k}$.
\par The infrared divergence index of each term in this sum can be
cast in the form
\begin{equation}
    \omega_{ir}^{G_i} = \omega_{ir} - \widetilde{\omega}_{ir} +
    \widetilde{\omega}'_{ir}
    \nonumber
\end{equation}
where $\omega_{ir}$ is the divergence index of the unrenormalized
diagram $G$, $\widetilde{\omega}_{ir}$ is a contribution of the
unrenormalized subgraph $G_i$ to the infrared divergence index
$\omega_{ir}$ and $\widetilde{\omega}'_{ir}$ is the contribution of
$M(G_i)F_i$.
\par In the unrenormalized diagram (Fig.~\ref{fig_large})
propagators at the boundary of the diagram do not contribute to
$\omega_{ir}$, because the external momenta are nonexceptional.
Let us pull all these propagators together to one point, then we
obtain a vertex with $n$ external and $n_{int}$ internal lines.
Using the identity, which relates the number of loops to the
number of vertices and internal lines, as well as the topological
identity, which connects the number of internal and external lines
with the number of vertices of various types, we find
$\omega_{ir}=n_{int}-2$. As we are dealing with
one-particle-irreducible graphs, $n_{int}\geq 2$, and thus
$\omega_{ir}\geq 0$.
\par Let us evaluate $\widetilde{\omega}'_{ir}$ for some subgraph.
The first term in the Taylor expansion is the function $F_i$ at the
point $\widetilde{k}$ (by $\widetilde{k}$ we denote the set of all
external momenta of the graph $G_i$), i.e. we need to calculate the
IR divergence index of the graph $G_i$ with fixed external momenta
$\widetilde{k}$. It can be done in a fashion analogous to the way we
calculated $\omega_{ir}$ for the unrenormalized diagram. For a
subgraph with no mass insertion $\widetilde{\omega}'_{ir}\geq 2$ at
nonexceptional $\widetilde{k}$. As for $\widetilde{\omega}_{ir}$,
$\widetilde{\omega}_{ir} = 4\widetilde{L} + \widetilde{V}_3 -
2\widetilde{I}$, where $\widetilde{L}$, $\widetilde{V}_3$,
$\widetilde{I}$ are the numbers of loops, triple vertices and
propagators in ~$G_i$, respectively. We see that
$\widetilde{\omega}_{ir}$ coincides with the UV divergence index of
the subgraph $G_i$, and $\omega_{uv}\leq 2$ ($\omega_{uv}=4-E$,
where $E$ is the number of external lines; $\omega_{uv}=2$ is
saturated only for self-energetic diagrams). Thus, taking into
account that $\widetilde{\omega}'_{ir}\geq 2$,
$\widetilde{\omega}_{ir}\leq 2$ and $\omega_{ir}\geq 0$, we obtain:
\begin{equation}\label{omega_ir_Gi}
    \omega_{ir}^{G_i} = \omega_{ir} + ( \widetilde{\omega}'_{ir} -
    \widetilde{\omega}_{ir} ) \geq \omega_{ir} \geq 0
\end{equation}
In the next terms of the Taylor expansion differentiation acts on
the part of the diagram $G_i$, through which the fixed momentum
$\widetilde{k}$ flows. This part of the subgraph $G_i$ does not
influence the index $\widetilde{\omega}'_{ir}$, and the relation
(\ref{omega_ir_Gi}) still holds. If $G_i$ includes a mass
insertion, then $\widetilde{\omega}'_{ir}\geq 0$, but also
$\widetilde{\omega}_{ir}\leq 0$. Indeed, it is easy to verify that
$\widetilde{\omega}_{ir} = 2-E$ (where $E$ is the number of
external lines of $G_i$). For 1PI diagrams $E\geq 2$, therefore
$\widetilde{\omega}_{ir}\leq 0$. Thus, $\omega_{ir}^{G_i}\geq 0$
also in the case when $G_i$ includes a mass insertion. Note that
the normalization points $\widetilde{k}$ should be chosen
nonsingular. For example, the zero subtraction points should be
excluded form the very beginning, because from the arguments
described above it is clear that in this case
$\widetilde{\omega}'_{ir}$ does not satisfy inequality
$\widetilde{\omega}'_{ir}\geq 2$ as the zero momenta are
exceptional.

We have proved that the divergence at zero external momentum is
not worse than logarithmic, and in this case the limit (\ref{17})
is zero, and (\ref{15}) is true.

Suppose that at some value of the gauge fixing parameter $\alpha$
the condition $\langle \tr A_{\mu}^{2} \rangle \neq 0$ is
fulfilled, then we define an operator
\begin{equation}\label{20}
D  \overset{def}{=} \underset{\Omega \rightarrow \infty}{{\rm
lim}} \left( \frac{1}{\Omega} \cdot \frac{ \int_{\Omega} \,{\rm
dx}\,{\rm tr} (A_{\mu}^{2})}{\langle {\rm tr} (A_{\mu}^{2})
\rangle} \right)
\end{equation}
It was shown above that when we consider a Green's function of a
few elementary fields and an operator $D$ we can substitute the
latter with a unit operator (see (\ref{15})). In spite of this,
the operator $D$ is not identical to the unit operator. Indeed,
Green's functions of operator $D$ and some other composite
operator (for example, $\tr(A_{\mu}^{2}(y))$), requires additional
renormalization, which is not connected to the renormalizations of
the separate composite operators $\tr (A_{\mu}^{2})$ entering the
Green's function.

\section{Discussion}

In this work the following results have been obtained: first of
all, it has been shown that in noncommutative field theory the
generalized Ward identities hold for Green's function involving
operator $\int \,{\rm dx}\,{\rm tr}(A_{\mu}^{2}-\frac{2}{g \xi}
x^{\nu} \theta_{\mu\nu}A^{\mu})$. In commutative field theory Ward
identities for Green's functions with an insertion of an operator
$D$ (see (\ref{20})) hold as if it were a unit operator. It has
been shown that in noncommutative theory classical gauge
invariance and generalized Ward identities do not guarantee gauge
independence of the matrix elements of operators, corresponding to
functions that are invariant under classical gauge
transformations. It is due to the fact that there are no local
gauge invariants in noncommutative theories. An analogous
situation is encountered, for example, in gravity theory and in
supersymmetric theories. It would be interesting to explore the
possible physical consequences of this fact.

As it was discussed above, the use of the condensate $\langle \tr
(A_{\mu}^{2})\rangle$ as an order parameter is only possible if
the collection of values of the coupling constant, at which it
vanishes, is gauge independent. Note that the minimal value of
$\langle \tr (A_{\mu}^{2})\rangle$ with respect to gauge
transformations, which was proposed in the paper ~\cite{zakharov}
as a candidate for an order parameter, fulfills this condition.
The question, whether this requirement is satisfied for the
condensate $\langle \tr (A_{\mu}^{2})\rangle$, is still open.
\newline

\textbf{Acknowledgements.} This work has been partially supported
by grants RFBR № 050100541, the grant for the support of leading
scientific schools № 20052.2003.1 and the RAS program "Theoretical
problems of mathematics".

\end{document}